\documentclass{elsart5p}
\usepackage{graphicx,natbib,amssymb}
\journal{New Astronomy}
\newcommand{\cmt}{\,cm$^{-3}$}

\newcommand{\ecsa}{$\rm\,erg\,cm^{-2}\,s^{-1}\,\AA^{-1}$}
\newcommand{\ha}{H$\alpha$}
\def\I{\rm {\scriptsize I}}
\begin{document}
\begin{frontmatter}

\title{On the effect of emission lines on the $UBVR$ photometry}

\author{A. Skopal\thanksref{fn1}}

\thanks[fn1]{E-mail: skopal@ta3.sk}

\thanks[fn2]{doi:10.1016/j.newast.2007.04.003}

\address{Astronomical Institute, Slovak Academy of Sciences,
         059\,60 Tatransk\'{a} Lomnica, Slovakia \\[2mm]
    {\rm Received 2 March 2007; accepted 9 April 2007}\thanksref{fn2}}

\begin{abstract}
 We investigate the effect on the $U,~B,~V,~R_{\rm C}$ and 
  $R_{\rm J}$ magnitudes of the removal of emission lines 
  from a spectrum. 
  We determined $\Delta m$ corrections from the ratio 
   of fluxes with and without emission lines, transmitted from 
   the object through the photometric filter. An exact and 
   a simplified approach for operative use were applied. 
  The effect was demonstrated for classical symbiotic stars, 
   symbiotic novae and the classical nova V1974\,Cyg. It was 
   found that about 20-30\%, 30-40\%, 10\% and 26/20\% of 
   the observed flux in the $U$, $B$, $V$ and 
   $R_{\rm C}/R_{\rm J}$ filter, respectively, is radiated 
   in the emission lines of the investigated classical 
   symbiotic stars. 
   The largest effect was found 
   for symbiotic novae (RR\,Tel and V1016\,Cyg) and the classical 
   nova V1974\,Cyg at 210 days (in average of 74\%, 79\%, 56\% 
   and 66/60\%), because of their very strong emission 
   line spectrum. In all cases the line corrected flux-points 
   fit well the theoretical continuum. 
   The difference between $\Delta m$ corrections obtained by the 
   accurate calculation and that given by our approximate formula 
   is less than 10\%. Deviations up to 30\% can be only in the $U$ 
   passband. Examples for practical application are suggested. 
\end{abstract} 
\end{frontmatter}
%
%

\section{Introduction}

In many astrophysical applications photometric measurements 
through the standard $U,~B,~V,~R$ filters are used to analyze 
radiation in the continuum from stellar objects. 
For example, a diagnostic by the ($U-B,~B-V$)-diagram is 
frequently applied to compare the observed colour indices 
to those of the continuum radiation. 
%
%
However, photometric magnitudes represent integrated fluxes 
that include both the continuum and the line spectrum. 
Therefore, a correction for lines has to be applied 
to obtain photometric flux-points of the true continuum. 

The effect on the $U,~B,~V$ magnitudes of the removal of 
absorption lines from the spectrum was already introduced by 
\citet{seg59}. They found that the absorption lines cause 
{\em fainter} magnitudes at all wavelengths. 
On the other hand the presence of emission lines in the spectral 
region of the photometric passbands leads to {\em brighter} 
magnitudes than those of the continuum. However, it is difficult 
to quantify this effect. A strong variation of the emission 
spectrum (e.g. due to outbursts and/or orbital motion), large 
differences between individual objects and the complex profile 
of the true continuum of some interacting binaries (e.g. 
symbiotic stars, classical novae) preclude a simple solution. 
%
%
Therefore this problem has been approached only individually 
and without giving a concept for a general application. 
Probably a first more thorough approach was described by 
\citet{m+82} who corrected their $V$ magnitudes for strong 
emission lines in the spectrum of AR\,Pav. 
Other authors briefly reported just their results on the 
emission lines effect to evaluate the continuum level in the 
optical region \citep[e.g.][]{f-c+95,ttt03}. Recently \citet{sk03} 
suggested an accurate calculation of $\Delta m$ corrections 
for the $UBV$ passbands. 

In this paper we aim to quantify the effect of emission lines 
on the $U,~B,~V,~R_{\rm C},~R_{\rm J}$ magnitudes by the exact 
approach and to derive an approximate formula for an operative 
use. In Sect.~2 we introduce our method, in Sect.~3 demonstrate 
the effect on a sample of selected emission-line objects and 
in Sect.~4 we briefly discuss our results and suggest their 
practical application. 
%
%
\begin{figure}
\centering
\begin{center}
\resizebox{\hsize}{!}{\includegraphics[angle=-90]{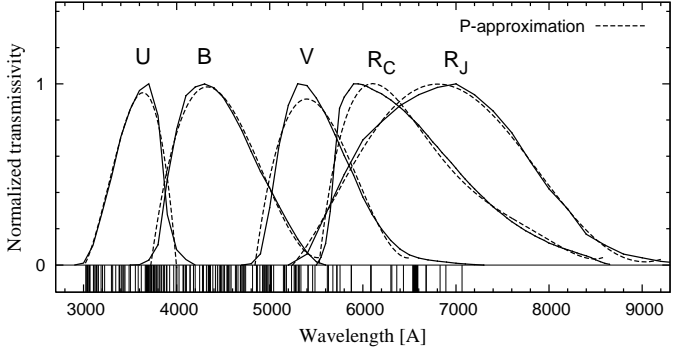}}
\caption[]{
Response functions for the $U,~B,~V,~R_{\rm C}$ and $R_{\rm J}$ 
filters after \citet{ms63}, \citet{b79} and \citet{johnson}, 
respectively. Dashed lines represent their polynomial 
approximation (Table~1). The bottom band shows distribution 
of the emission lines from the spectrum of RR\,Tel (Sect.~3.3.2). 
           }
\end{center}
\end{figure}

\section{The method}

\subsection{An exact approach}

The aim of this paper requires the ratio of the continuum with 
the superposed emission lines to the line-removed continuum at 
all relevant wavelengths. Therefore, for the purpose of this 
paper we express the observed flux in the form 
%
%
\begin{equation}
  F_{\rm obs}(\lambda) = 
  F_{\rm cont}(\lambda) (1 + \epsilon (\lambda)), 
\end{equation}
%
where $F_{\rm cont}(\lambda)$ is the true continuum (i.e. 
line-removed continuum) and $\epsilon (\lambda)$ represents 
the emission line spectrum in units of the continuum at the 
wavelength $\lambda$. Then the magnitude difference, 
$\Delta m$, between the observed magnitude, $m_{\rm obs}$, and 
that corresponding to the line-removed continuum, 
$m_{\rm cont}$, can be expressed as 
%
%
\begin{eqnarray}
\Delta m = m_{\rm obs} - m_{\rm cont} = 
~~~~~~~~~~~~~~~~~~~~~~~~~~~~~
\nonumber \\
-2.5\log\,\biggl[\frac{\int_{\lambda}\!\! F_{\rm cont}(\lambda)S(\lambda)
             (1 + \epsilon (\lambda))\,{\rm d}\lambda}
             {\int_{\lambda}\!\!F_{\rm cont}(\lambda)S(\lambda)\,
               {\rm d}\lambda} \biggr],
\end{eqnarray}
%
where $S(\lambda)$ are response functions for the given filter 
(Fig.~1). We approximate the emission spectrum with an ensemble 
of Gaussian functions, $G_{\rm i}$, as 
%
\begin{equation}
 \epsilon (\lambda) = \sum_{i}G_{\rm i}(\lambda;\lambda_{\rm i},
                      I_{\rm i},\sigma_{\rm i}), 
\end{equation}
%
where $\lambda_{\rm i}$ is the wavelength of the i-$th$ line, 
$I_{\rm i}$ its maximum in units of the local continuum and 
$\sigma_{\rm i}$ its HWHM. For the purpose of this work we 
express the Gauss function in the form 
%
\begin{equation}
 G_{\rm i}(\lambda;\lambda_{\rm i},I_{\rm i},\sigma_{\rm i}) = 
          I_{\rm i}\times \exp\bigg[-\frac{1}{2}\bigg(\frac{\lambda -
          \lambda_{\rm i}}{\sigma_{\rm i}}\bigg)^2\bigg].
\end{equation}
%
According to Eq.~(2) the removal of emission lines from the 
spectrum gives {\em fainter} magnitudes at all wavelengths. 
Exact calculation of $\Delta m$ corrections according to 
Eq.~(2) is complicated with modeling the true continuum 
(Sect.~3). Therefore, for the practical use, we suggest 
an approximate solution. 

\subsection{A simplified approach}

In this section we simplify calculation of Eq.~(2) with 
the following assumptions:
%

(i) A constant level of the continuum, $F_{\rm cont}(\lambda)$, 
within the photometric passbands. 

(ii) Approximation of the $S(\lambda)$ functions  by 
a polynomial function (Fig.~1, Table~1). 

(iii) The response function $S(\lambda)$ is constant within the 
line, because emission lines are relatively narrow with respect 
to the width of the photometric filter. 

(iv) According to the relation (4), we express the flux of 
the i-$th$ line in the continuum units as 
%
\begin{equation}
  F_{\rm i} = 
  \int_{\lambda}G_{\rm i}(\lambda;\lambda_{\rm i},I_{\rm i},
  \sigma_{\rm i})\,{\rm d}\lambda\,=\,\sqrt{2\pi}\,
  I_{\rm i}\sigma_{\rm i}. 
\end{equation}
%
%
Under these assumptions Eq.~(2) can be expressed in the form 
%
\begin{equation}
\Delta m_f =
  -2.5\log\biggl[1 + \frac{\sqrt{2\pi}}{C_{f}}\sum_{\rm i}
   P_{f}(\lambda_{\rm i})I_{\rm i}\sigma_{\rm i}\biggr],
\end{equation}
where the index $f$ denotes the considered filter 
and $P_{f}(\lambda_{\rm i})$ is the filter transmissivity 
at the wavelength $\lambda_{\rm i}$ of the i-$th$ line 
given by the polynomial approximation of the corresponding 
$S(\lambda)$ function, and 
%
\begin{equation}
 C_{f}\,=\,
   \int_{\lambda}\!S_{f}(\lambda)\,{\rm d}\lambda. 
\end{equation}
%
The polynomial coefficients and the  constants $C_{f}$ 
are in Table~1. Both are determined for wavelengths in \AA. 
According to Eqs.~(2) and (6), the $\Delta m$ corrections depend 
in great deal on the ratio of the line fluxes to the level 
of the local continuum and their position $\lambda_{\rm i}$ 
(i.e. $P_{f}(\lambda_{\rm i})$). 
To derive appropriate $\Delta m$ corrections the spectrum 
can be in arbitrary units (Eq.~(2)). 
%
%
%
\begin{table}
\begin{center}
\caption{
Polynomial approximation of the transmission functions 
$S(\lambda) \equiv P(\lambda)$ (Fig.~1) and the constants 
$C_{f}$ (Eq.~7). 
        }
\begin{tabular}{cccccc}
\hline
\hline
       &\multicolumn{5}{c}{
        $P(\lambda)=a_0+a_1\lambda+a_2\lambda^2+a_3\lambda^3$~
        (+$a_4\lambda^4+a_5\lambda^5$)}                       \\
$f$    &  $a_0$ &  $a_1$  &   $a_2$   &   $a_3$   &$C_{f}$\\
\hline
 U     & 169.89 &-0.15959 & 4.9442E-5 &-5.0407E-9 & 567   \\
 B     &-127.02 & 0.08009 &-1.6508E-5 & 1.11763E-9 & 1017   \\
 V     &-276.23 & 0.14212 &-2.4120E-5 & 1.35310E-9 & 876  \\
R$_{\rm C}$ 
       &-2425.1 & 1.65606&-4.50068E-4 &6.09087E-8 & 1452   \\
       &\multicolumn{5}{c}{$a_4$=-4.10712E-12,~
                           $a_5$=1.104315E-16}             \\
R$_{\rm J}$ 
       &345.733 &-0.24638&6.8274e-05&-9.1966e-09& 2070   \\
       &\multicolumn{5}{c}{$a_4$=6.03503e-13,~
                           $a_5$=-1.54735e-17}             \\
%
\hline
\end{tabular}
\end{center}
\end{table}

\subsection{Relative contributions from lines}

To determine the flux emitted in lines, $F_{l}$, relative 
to the total observed flux, $F_{\rm obs}$, throughout the given 
filter (i.e. the $F_{l}/F_{\rm obs}$ ratio), we rewrite 
Eq.~(2) in the form 
%
\begin{equation}
   \Delta m = -2.5\log\,\bigl[F_{\rm obs}/F_{\rm cont}\bigr]. 
\end{equation}
As $F_{\rm obs} = F_{\rm cont} + F_{l}$, the relative 
amount of radiation emitted in lines can be expressed as 
%
\begin{equation}
  F_{l} / F_{\rm obs} = 1 - 10^{0.4\,\Delta m} .
\end{equation}
Table~2 introduces these ratios for investigated objects 
in percents. 

\section{Analysis and results}

To demonstrate the effect of emission lines on the $UBVR$ 
photometric measurements we selected four groups of objects 
characterized with similar line spectrum and a complex profile 
of the continuum. They are: 

(i) Classical symbiotic stars that display typical quantities 
of $F_{\rm cont}(\lambda)$ and $I_{\rm i}$, mainly during their 
quiescent phases. We present examples of 
AX\,Per, AR\,Pav, AG\,Peg and Z\,And. 

(ii) Objects with a low level of the continuum resulting in 
higher values of relative line fluxes. We represent this case 
by the AX\,Per spectrum during its 1994 total eclipse. 

(iii) Objects with very rich and intense emission line spectrum. 
Here we selected examples of two symbiotic novae, V1016\,Cyg and 
RR\,Tel. 

(iv) Objects with extremely broad and intense lines that usually 
develop during nebular phases of classical novae. We give 
example of the V1974\,Cyg nova at 210 days. 

Following Eqs.~(1) and (2) we need the profile of the continuum 
and the emission line spectrum. 
%
%
We reconstructed the true continuum, $F_{\rm cont}(\lambda)$, 
according to \citet{sk05} (hereafter S05). 
This allow us to compare directly the line-corrected $UBVR$ 
flux-points with the theoretical continuum (Figs.~2, 4--6 below). 
%
%
The emission line spectrum, $\epsilon(\lambda)$, was 
reconstructed according to relations (3), (4) and (5) with 
the aid of its parameters available in the literature 
(mostly fluxes, see below). 
%
%
The $U,~B,~V,~R$ magnitudes were converted to fluxes according 
to the calibration of \citet{hk82} and \citet{b79}. 
Simultaneous observations are required, because of a strong
variation in the line spectrum and the continuum, mainly in 
the case of interacting binaries. 
All observations were dereddened for interstellar extinction. 
$E_{\rm B-V}$ quantities were taken in most cases from 
Table~1 of S05. 
%

In the following sections we introduce above mentioned 
examples. Resulting $\Delta m$ corrections due to emission 
lines are listed in Table~2. 

\subsection{Classical symbiotic stars} 
%
%

\subsubsection{AX\,Per}

AX\,Per is the symbiotic binary with a high orbital 
inclination. During active phases its light curve shows narrow 
minima -- eclipses -- that change into pronounced wave-like 
orbitally-related variation during quiescence 
\citep[e.g.][]{sk+01}. 

As the true continuum $F_{\rm cont}(\lambda)$ we adopted the 
model of S05 made for ultraviolet observations from 07/11/93 
($\varphi = 0.57$) taken during the post-outburst activity. 
Figure~2\,(a) shows a detail covering the near-UV/optical region. 
%
To reconstruct the function $\epsilon(\lambda)$ we used emission 
lines published by \citet{ss40} and \citet{lb54,lb57} observed 
on 1952 October and 1956 August ($\varphi = 0.5 - 0.6$). They 
were obtained at the same orbital phase and the level of 
the activity as those on 07/11/93 
\citep[cf. Figs.~2 and 3 of][]{sk+01}. 
The line intensities were re-scaled with respect to the local 
continuum according to corresponding intensity tracings presented 
in the figures of the above referred papers. Quantities 
of $\sigma_{\rm i}$ were adopted to 1$-$2\,\AA\ for hydrogen 
lines and 0.5$-$1\,\AA\ for other lines. 
%
The $U,~B,~V$ measurements were taken simultaneously to the IUE 
spectra \citep{sk+01}. The available magnitude $R_{\rm J}$ = 9.83 
was measured on JD~2\,446\,332 (23/09/85, $\varphi$ = 0.2) 
\citep{m+92}. 

The removal of emission lines makes the star's brightness fainter 
by 0.43, 0.42, 0.10, 0.17 and 0.11\,mag in the $U$, $B$, $V$, 
$R_{\rm C}$ and $R_{\rm J}$ filter, respectively. The corrected 
$UBVR_J$ measurements are close to the predicted continuum 
(Fig.~2\,a). 
%
%
%
\begin{figure*}
\centering
\begin{center}
\resizebox{\hsize}{!}{\includegraphics[angle=-90]{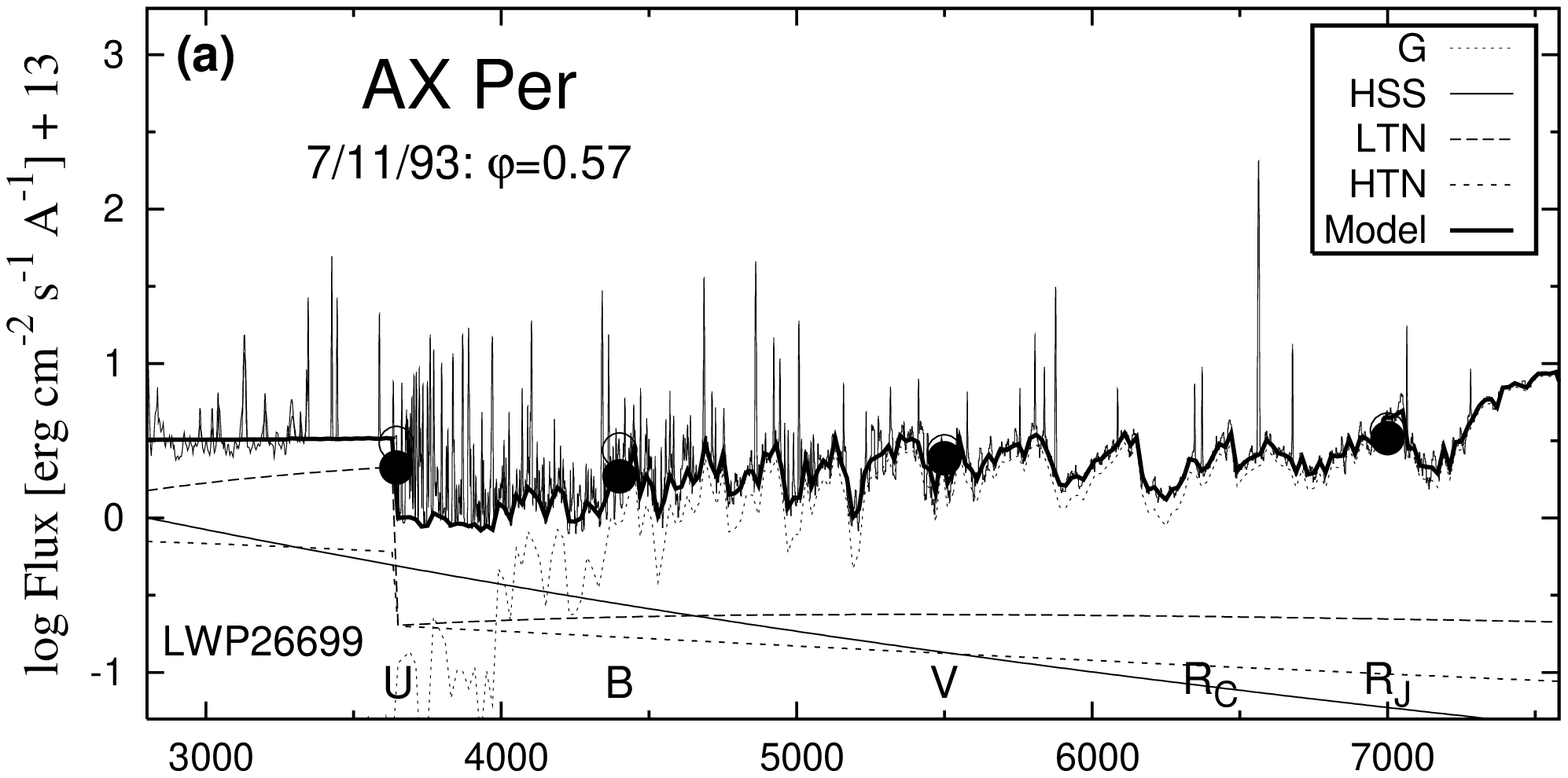}
                      \includegraphics[angle=-90]{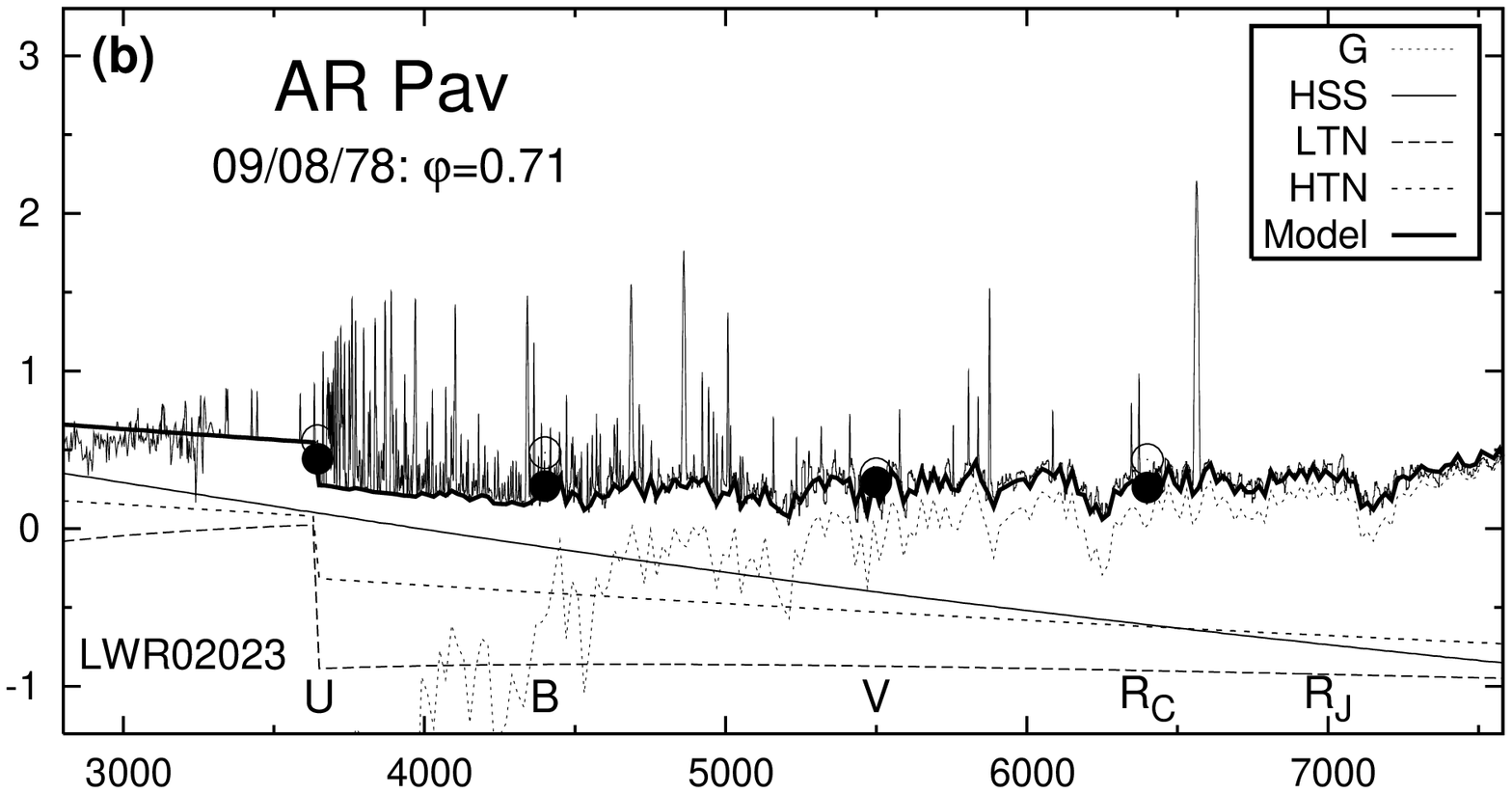}}
\resizebox{\hsize}{!}{\includegraphics[angle=-90]{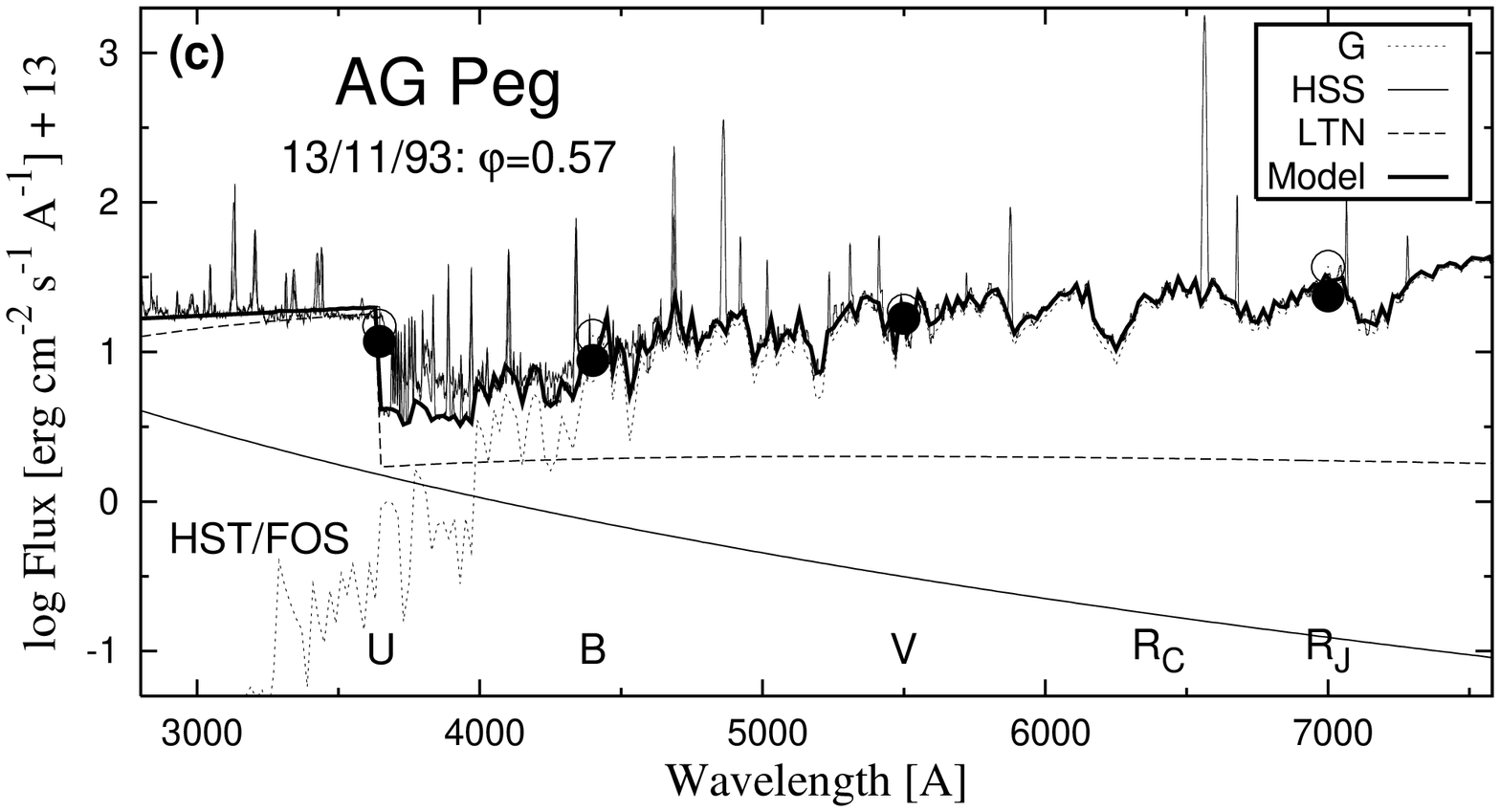}
                      \includegraphics[angle=-90]{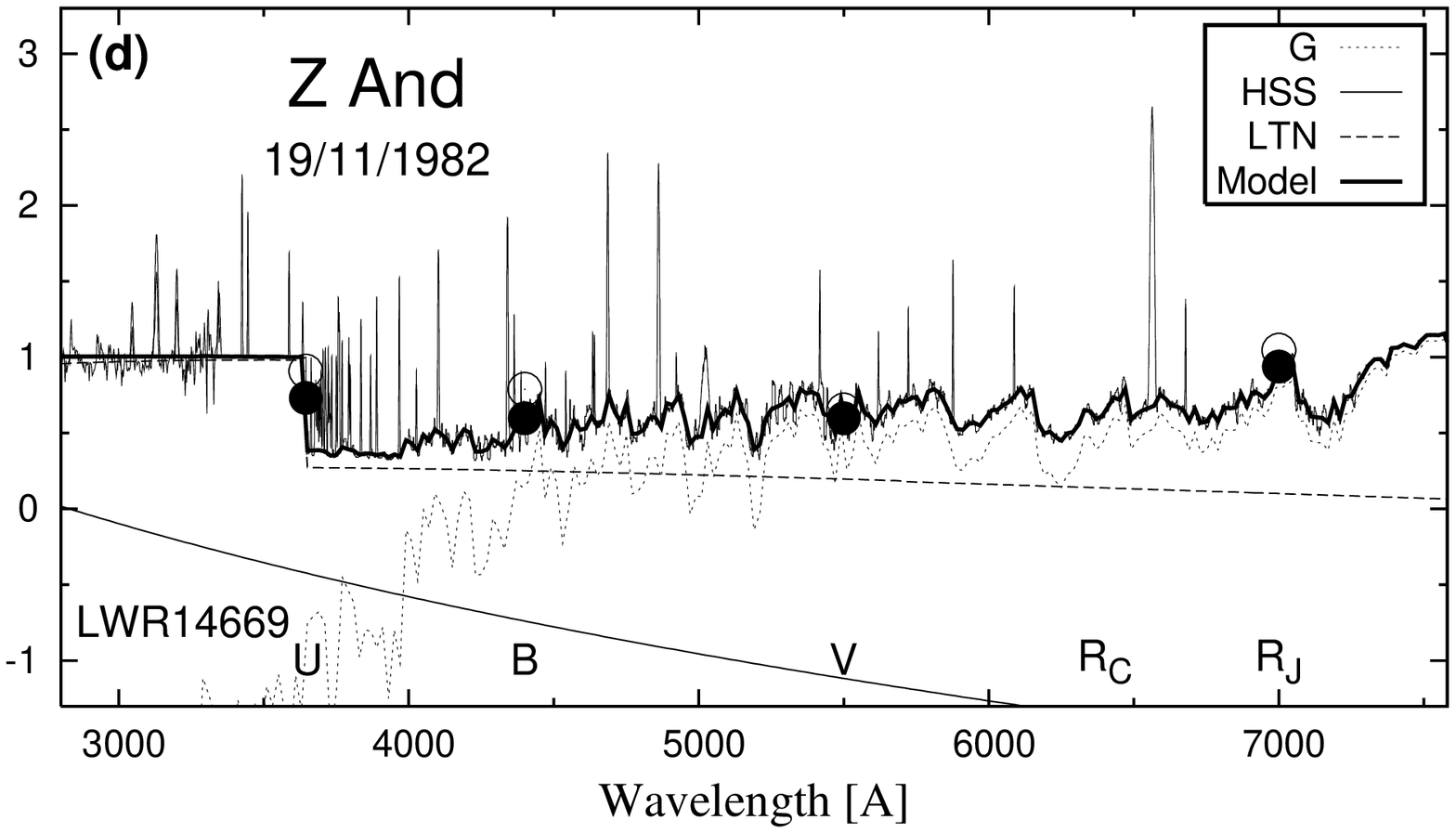}}
\caption[]{
Reconstructed SEDs for the investigated classical symbiotic 
stars in the near-UV/optical region. Individual components of 
radiation are denoted as in S05: 
   HSS (hot stellar source), 
   LTN (low-temperature nebula), 
   HTN (high-temperature nebula) and 
    G (radiation from the red giant). 
The model (solid thick line = $F_{\rm cont}(\lambda)$ in 
Eq.~(2)) is given by their superposition. Emission line 
spectrum, $\epsilon(\lambda)$ (Eq.~3), is superposed on the 
model. Open and full circles are flux-points of the observed 
and corrected $U,~B,~V,~R_{\rm C},~R_{\rm J}$ magnitudes, 
respectively. 
}
\end{center}
\end{figure*}
%
%

\subsubsection{AR\,Pav}

AR\,Pav is another eclipsing symbiotic star. The permanent 
presence of deep and narrow minima in its historical light 
curve, strong out-of-eclipse variations \citep[e.g.][]{bruch} 
and the observed symbiotic phenomenon, suggest that AR\,Pav 
persists in an active phase (S05). 

We used the IUE spectra SWP02236 and LWR02023 exposed on 09/08/78 
($\varphi$ = 0.71) to model the true continuum (Fig.~2\,b). It is 
very similar to that from 10/05/81 (S05). 
%
The line spectrum of AR\,Pav was reconstructed in part on 
the basis of observations published by \citet{th74}, who 
presented parameters of the strongest emission lines. 
We note that during active phases of symbiotic stars, 
profiles of H\,\I\ and He\,\I\I\ emission lines 
are significantly broader than during quiescence 
\citep[e.g.][]{th74,sk06}. 
In addition, the function $\epsilon(\lambda)$ was complemented 
with that for AX\,Per, because their UV spectra are very 
similar (Fig.~3). 
%
%
There is no simultaneous photometric measurements to 
the spectroscopic observations. 
Therefore we used average $U,~B,~V$ magnitudes from 
\citet{a74} and scaled them so to fit the end of the LWR 
spectrum. The magnitude $R_{\rm C} = 10.1$ was taken from 
\citet{m+82} obtained through their red channel and out of 
the eclipse. 

Also here the largest effect is in the $B$ band (-0.55\,mag) 
and smallest in $V$ (-0.13\,mag). The same value of 
$\Delta V$ = -0.13\,mag was derived also by \citet{m+82}. 
The line corrected fluxes fit 
well the theoretical continuum (Fig.~3\,b). 
%
%

\subsubsection{AG\,Peg}

AG\,Peg is known as the slowest symbiotic nova with eruption 
in the mid-1850's. Currently it displays all signatures of 
a classical symbiotic star. 
%
%
Lines of H\,\I\ and He\,\I\ vary as a function of 
the orbital phase with a trend following a slow fading of 
the star's brightness \citep{k+93}. In this case it is 
important to analyze observations taken at the same orbital 
phase during the same or neighbouring cycles. 

As the true continuum we adopted the model made by S05 
on the basis of the HST/FOS observations from 13/11/93 
($\varphi$ = 0.6, i.e. around the light maximum). 
Figure~2\,(c) shows its near-UV/optical part. 
%
The function $\epsilon(\lambda)$ was reconstructed according 
to the line spectrum of the FOS observations to 4780\,\AA. 
From this wavelength to 7300\,\AA\ we used the line fluxes 
from the maxima in Table~6 of \citet{k+93}. 
Quantities of $I_{\rm i}$ and $\sigma_{\rm i}$ were obtained 
from the fluxes and the level of the local continuum according 
to their Fig.~1 to satisfy Eq.~(5). 

The $U,~B,~V$ measurements were taken from the maximum around 
JD\,2\,450\,000 according to observations of \citet{tt98}. 
The adopted magnitude $R_{\rm J}$ = 6.92 was measured around 
the 1973 maximum \citep{f85}. 
%

For AG\,Peg the true continuum is fainter by 0.19, 0.41, 0.12, 
0.40 and 0.31\,mag than the star's brightness measured directly 
through the $U,~B,~V,~R_{\rm C}$ and $R_{\rm J}$ filters, 
respectively. 
Figure~2\,(c) shows that the corrected fluxes are close to the 
predicted continuum. 
In spite of a relatively high level of the red continuum, 
the influence of lines to the $R$-continuum is significant, 
because of a strong \ha\ emission. 
%

\subsubsection{Z\,And}

Z\,And is a prototype of the class of symbiotic stars. Its 
historical light curve (from 1887) shows phases of activity 
with up to 2-3\,mag increase of the star's brightness, 
alternating with periods of quiescence. The quiescent phase 
is characterized by a complex wave-like light variation as 
a function of the orbital phase \citep[e.g.][]{fl94}. 

As an example of the true continuum we adopted the model of 
S05 made for 19/11/82 ($\varphi$ = 0.49). Figure~2\,(d) shows 
its near-UV/optical portion. 
%
The line spectrum was reconstructed according to observations 
published by \citet{f-c+95}. The relevant spectrum was obtained 
on 1986 July 12-13. Some weaker lines were added from 
\citet{mk96} taken on 18/10/86. In addition, the region of the 
higher members of Balmer lines was enriched with lines from 
the $\epsilon(\lambda)$ function of AX\,Per. 
%
Simultaneous $U,~B,~V$ measurements were taken from \citet{bel92}. 
The magnitude $R_{\rm J} = 8.69$ was estimated from Belyakina's 
Fig.~1. It corresponds to a maximum during quiescence. 

To get the level of the continuum in the $U,~B,~V,~R_{\rm C}$ 
and $R_{\rm J}$ passbands one have to add 0.30, 0.48, 0.12, 
0.39 and 0.27\,mag, respectively, to the observed values. 
The corrected fluxes are close to the predicted continuum. 
%
%
\subsection{Symbiotic binaries during eclipses}

Here we present example of AX\,Per during its 1994 eclipse, 
for which ultraviolet and optical spectra as well as 
photometric observations were available. 

The low-resolution spectra from the IUE archive 
(SWP52027 + LWP29094) taken on 04/09/94 ($\varphi$ = 0.01) 
were used to model the true continuum. 
It is as faint as $\sim 5-6\times 10^{-14}$\ecsa\ and 
flat in the profile ($T_{\rm e}$(HTN)$\,\sim\,32\,000$\,K) 
in the UV with a dominant contribution from the giant 
at longer wavelengths beyond the $U$ passband (Fig.~4). 
%
The emission line spectrum was taken from Table~4 of \citet{mk92} 
(observations at eclipses on JD\,2\,447\,493 and/or JD\,2\,444\,901) 
and Table~3 of \citet{sk+01} (observation on 05/09/92, 
$\varphi = 0.01$). A few lines were included from the near-UV 
of the LWP29094 spectrum. 
%
%
$UBV$ photometric measurements were taken simultaneously to 
the IUE observations \citep{sk+01}. 
%

The low level of the continuum produces high values of the 
relative line fluxes (Eq.~(5)). As a result the observed 
photometric flux-points are well above the predicted continuum. 
The corrected $B$ and $V$ magnitudes fit the continuum well. 
However, the calculated correction $\Delta U = 0.77$\,mag 
is not sufficient. A correction of $\Delta U \sim 1$\,mag 
is required. This discrepancy is probably caused by incomplete 
line spectrum within the $U$ passband we had available. 
Note that in the case of a low level of the continuum also 
faint emission lines can make a distinctive effect. 
%
%
%
%
%
%
\begin{figure}
\centering
\begin{center}
\resizebox{\hsize}{!}{\includegraphics[angle=-90]{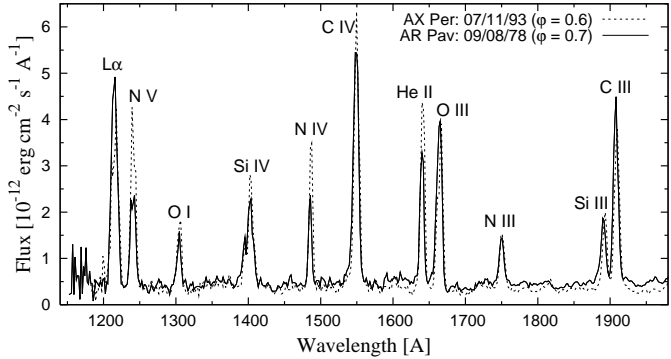}}
\caption[]{Close similarity of the ultraviolet spectra of 
           AX\,Per (07/11/93) and AR\,Pav (09/08/78). 
           The spectra are dereddened. 
           }
\end{center}
\end{figure}
%
%
%
\begin{figure}
\centering
\begin{center}
\resizebox{\hsize}{!}{\includegraphics[angle=-90]{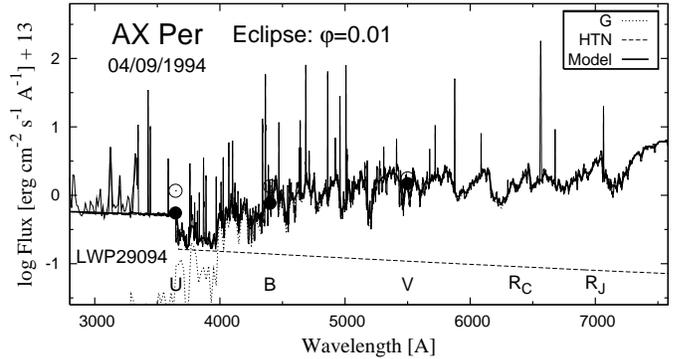}}
\caption[]{As in Fig.~2, but for AX\,Per during its total 
1994-eclipse. 
           }
\end{center}
\end{figure}

\subsection{Symbiotic novae}

\subsubsection{V1016\,Cyg}

%
V1016\,Cyg underwent its nova-like outburst in 1964 when increased 
its brightness in the photographic region by about 5\,mag, from 
around 16 to a maximumm of $\sim$10.5 in 1971. Afterwards it has 
continued a very slow gradual decrease by less than 1 mag to date 
\citep[][ and references therein]{par+00}. The binary contains 
a Mira variable as the cool component with a strong IR dust 
emission observed at/beyond the K band \citep[e.g.][]{ts00}. 

We modeled the optical continuum of V1016\,Cyg using the IUE 
spectra SWP24655 and LWP04959 taken on 10/12/84 and the 
synthetic spectrum for the red giant with 
$T_{\rm eff} = 2\,700$\,K and $\log(g) = 0.5$ \citep{h+99} scaled 
to the $J$-band flux, which is assumed to be free of the dust 
emission (Fig.~5\,a). 
%
%
The line spectrum was reconstructed from emission line fluxes 
published by \citet{ss90}. The spectrum was taken on 15/11/87 
at the INT telescope. 
%
The $UBVR_{\rm J}$ and $J$ photometry was taken from 
\citet{m+92}, \citet{par+00} and \citet{ts00}. Observations 
were dereddened with $E_{\rm B-V}$ = 0.28 \citep{ss90}. 

We found that the removal of emission lines makes the star's 
brightness fainter by 1.23, 1.67, 1.19, 1.87 and 1.57\,mag in 
the $U,~B,~V,~R_{\rm C}$ and $R_{\rm J}$ band, respectively. 
The corrected photometric fluxes fit perfectly the model 
continuum. 
%
%
%
%
\begin{figure}
\centering
\begin{center}
\resizebox{\hsize}{!}{\includegraphics[angle=-90]{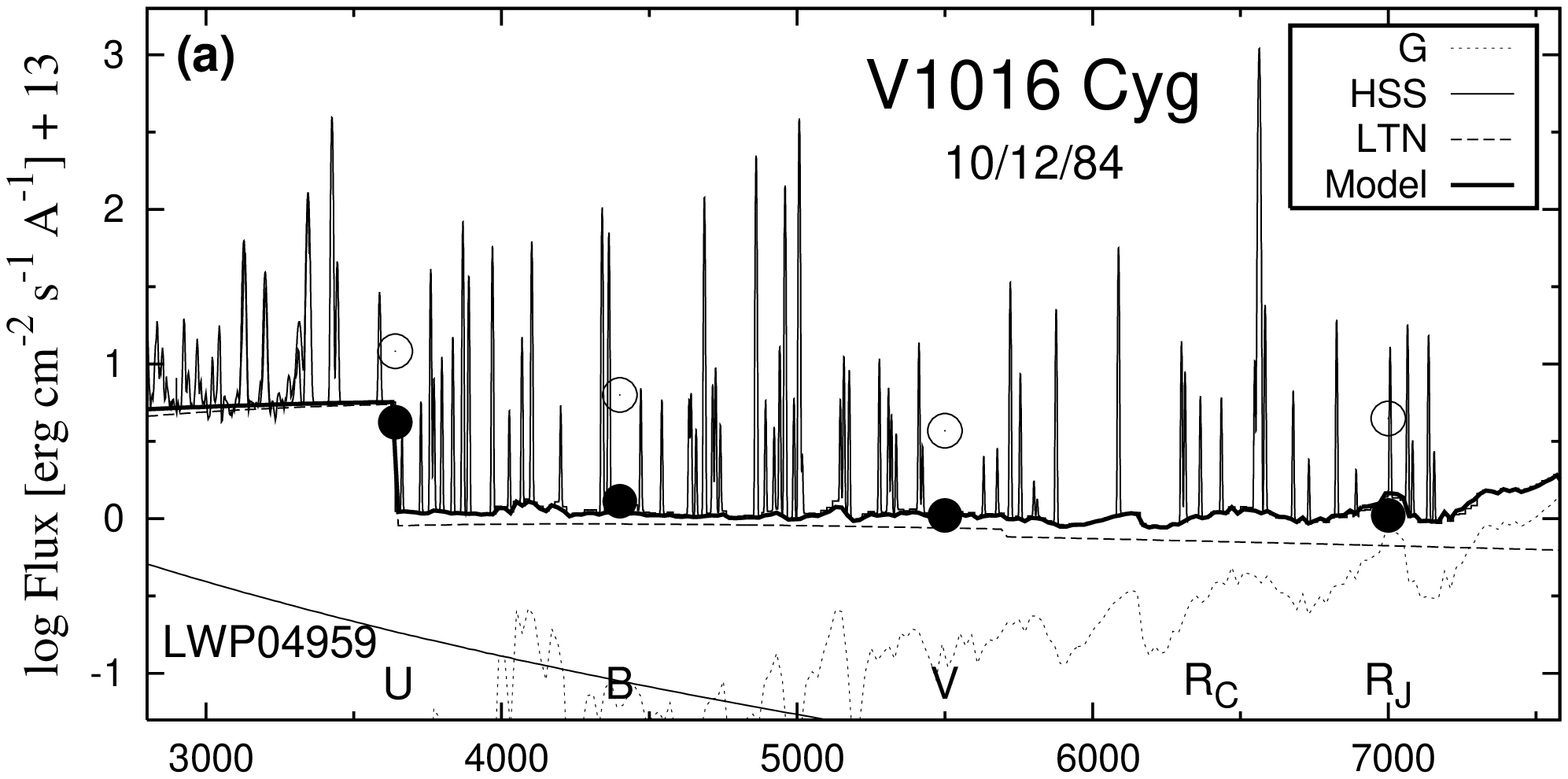}}
\resizebox{\hsize}{!}{\includegraphics[angle=-90]{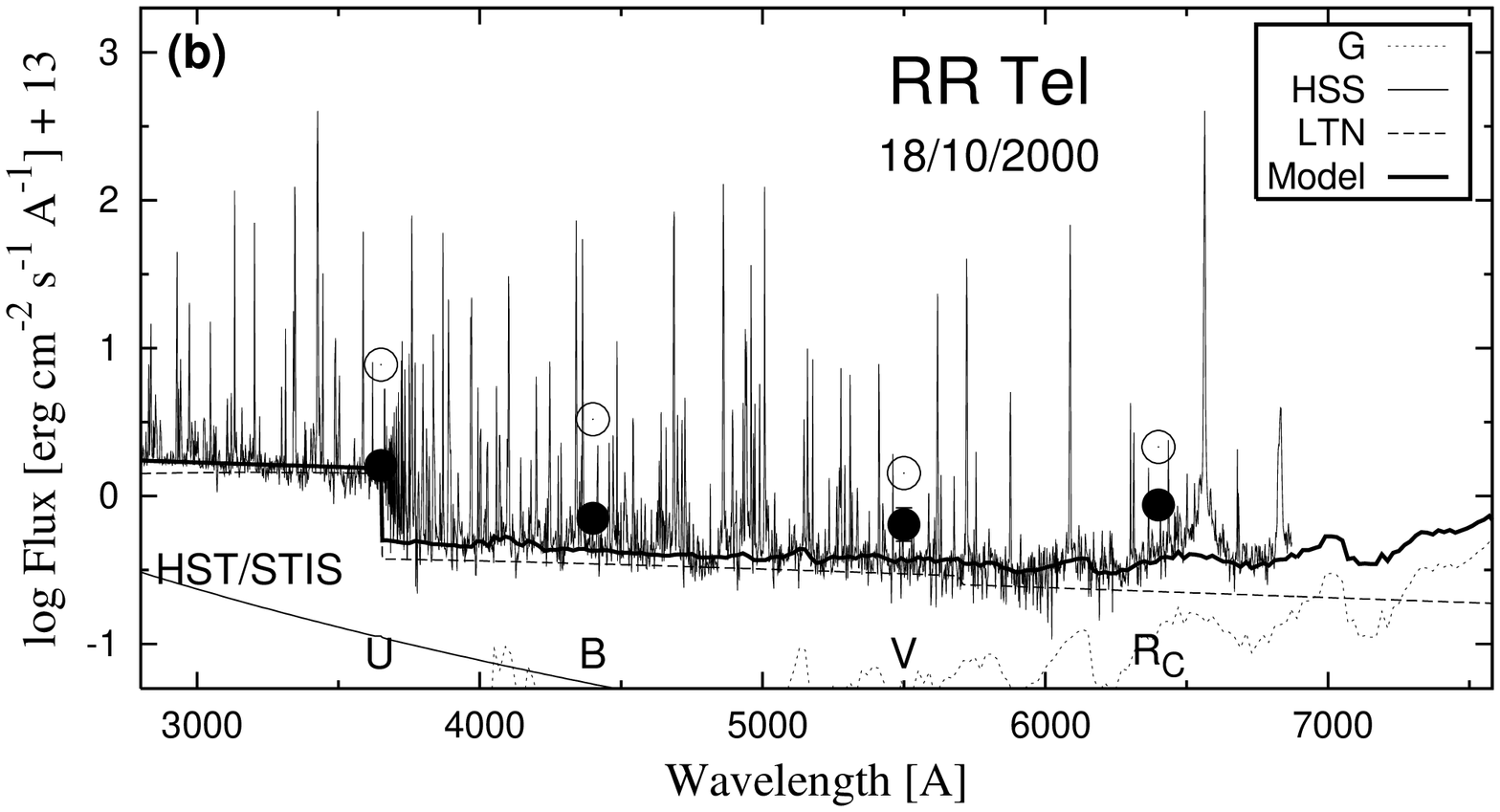}}
\caption[]{
As in Fig.~2, but for symbiotic novae V1016\,Cyg and RR\,Tel. 
           }
\end{center}
\end{figure}

\subsubsection{RR\,Tel}

%
RR\,Tel underwent a nova-like outburst in 1944 \citep[][]{p+83}. 
%
%
The light curve shows a gradual fading from the maximum 
with the visual magnitude around 11.4 during 2000-02 
\citep[see Fig.~1 of][]{kk+06}. The binary comprises a white 
dwarf as the hot component and a Mira-type variable as 
the cool component. Its circumstellar environment produces 
a strong nebular and dust emission \citep[e.g.][]{cf99}. 

We determined the true continuum by modeling the HST/STIS 
spectrum (1\,140 -- 7\,050\,\AA) taken on 18/10/2000 with 
the aid of the IR photometry published by \citet{kk+06}. The 
near-UV/optical region is very similar to that of V1016\,Cyg 
(Fig.~5). 
%
%
The emission line spectrum, $\epsilon (\lambda)$, was 
reconstructed from the observed line fluxes and the level 
of the local continuum. 
%
We used the $UBVR_{\rm C}$ magnitudes according to \citet{m+92}. 
They were obtained in 1990.5, i.e. about 10 years prior to 
the HST observation (2000.8), when the star was brighter by 
$\sim$\,0.7\,mag in the visual region \citep[see Fig.~1 of][]{kk+06}. 
Observations were dereddened for interstellar extinction with 
$E_{\rm B-V}$ = 0.10 \citep{y+05}. 

The influence of emission lines to the true continuum is 
significant. We found that RR\,Tel emitted about 79, 79, 55, 59 
and 49\% of the observed light in the emission lines throughout 
the $U,~B,~V,~R_{\rm C}$ and $R_{\rm J}$ filter, respectively. 
However, the flux-points of the line corrected magnitudes are 
still placed above the true continuum, because the photometric 
observations were carried out at a higher star's brightness 
than that corresponding to the HST observations. 
%
%
\begin{figure}
\centering
\begin{center}
\resizebox{\hsize}{!}{\includegraphics[angle=-90]{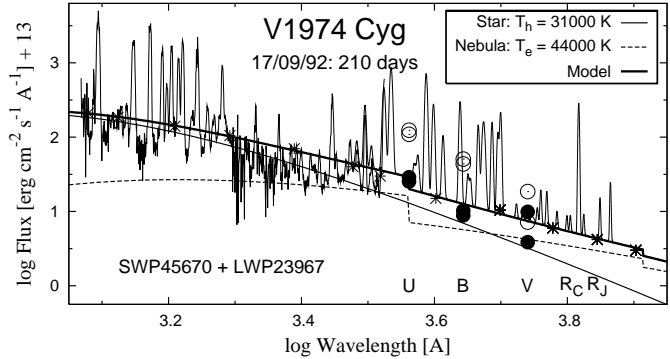}}
\caption[]{Spectrum of the classical nova V1974\,Cyg during 
           the nebular phase at 210 days. It was composed from 
           ultraviolet IUE and optical spectra of \citet{a+96}. 
           Continuum fluxes are denoted by ($\ast$). The model 
           SED (solid thick line) is described in the text. 
           Different magnitudes were obtained at different 
           observatories. 
           }
\end{center}
\end{figure}
%
%
\begin{table*}
\begin{center}
\caption{
Corrections $\Delta m(2)$ and $\Delta m(6)$ of 
the $UBVR_{\rm C}R_{\rm J}$ photometry for emission 
lines calculated according to Eqs.~(2) and (6), respectively. 
The relative amount of radiation emitted in lines and 
transmitted through the given filter 
(i.e. the $F_{l}/F_{\rm obs}$ ratio in Eq.~(9)) 
is in percents. 
        }
\begin{tabular}{lcccccccccc}
\hline
\hline
Object & 
\multicolumn{2}{c}{$\Delta U(2)/\Delta U(6)$}  &
\multicolumn{2}{c}{$\Delta B(2)/\Delta B(6)$}  &
\multicolumn{2}{c}{$\Delta V(2)/\Delta V(6)$}  &
\multicolumn{2}{c}{$\Delta R_{\rm C}(2)/\Delta R_{\rm C}(6)$} &
\multicolumn{2}{c}{$\Delta R_{\rm J}(2)/\Delta R_{\rm J}(6)$} \\
       &
[mag]  & [\%] &
[mag]  & [\%] &
[mag]  & [\%] &
[mag]  & [\%] &
[mag]  & [\%] \\
\hline
AG\,Peg      &
-0.19/-0.26  &~ 16/21~~ &
-0.41/-0.40  &~ 31/31~~ &
-0.12/-0.13  &~ 10/11~~ &
-0.40/-0.39  &~ 31/30~~ &
-0.31/-0.35  &~ 25/28 \\
Z\,And       &
-0.30/-0.35  &~ 24/28~~ &
-0.48/-0.48  &~ 36/36~~ &
-0.12/-0.14  &~ 10/12~~ &
-0.39/-0.44  &~ 30/33~~ &
-0.27/-0.38  &~ 22/30 \\
AR\,Pav      &
-0.40/-0.49  &~ 31/36~~ &
-0.55/-0.55  &~ 40/40~~ &
-0.13/-0.15  &~ 11/13~~ &
-0.37/-0.36  &~ 29/28~~ &
-0.29/-0.30  &~ 23/24 \\
AX\,Per      &
-0.43/-0.56  &~ 30/40~~ &
-0.42/-0.48  &~ 32/36~~ &
-0.10/-0.12  &~  9/10~~ &
-0.17/-0.20  &~ 14/17~~ &
-0.11/-0.16  &~ 10/14 \\
AX\,Per Ecl. &
-0.77/-0.60  &~ 51/42~~ &
-0.62/-0.64  &~ 44/45~~ &
-0.19/-0.33  &~ 16/26~~ &
-0.27/-0.32  &~ 22/26~~ &
-0.17/-0.26  &~ 14/21 \\
RR\,Tel      &
-1.71/-1.67  &~ 79/78~~ &
-1.68/-1.74  &~ 79/80~~ &
-0.87/-0.97  &~ 55/59~~ &
-0.98/-1.04  &~ 59/62~~ &
-0.73/-0.89  &~ 49/56 \\
V1016\,Cyg   &
-1.23/-1.15  &~ 68/65~~ &
-1.67/-1.71  &~ 78/79~~ &
-1.19/-1.33  &~ 67/71~~ &
-1.87/-1.61  &~ 82/77~~ &
-1.57/-1.46  &~ 76/74 \\
V1974\,Cyg   &
-1.57/-1.51  &~ 76/75~~ &
-1.73/-1.75  &~ 80/80~~ &
-0.66/-0.77  &~ 46/51~~ &
-0.94/-0.94  &~ 58/58~~ &
-0.88/-0.86  &~ 56/55 \\
\hline
\end{tabular}
\end{center}
\end{table*}

\subsection{Classical nova V1974~Cyg}

The classical nova V1974\,Cyg (Nova Cygni 1992) was discovered 
by \citet{c92} on 1992 February 19. It reached a peak visual 
magnitude of 4.4 on 1992 February 20.7 UT \citep{sch92}. 
V1974\,Cyg developed an emission line spectrum at early stages 
of its evolution. First nebular lines appeared in 1992 April 
and in 1992 September dominated the ultraviolet and optical 
spectrum \citep{ch+93,a+96}. To demonstrate the effect of emission 
lines on the $U,~B,~V$ measurements we used observations from 
the nebular phase taken in 1992 September. Observations were 
dereddened with $E_{\rm B-V}$ = 0.36 \citep{a+96}.

We reconstructed the continuum using the low-resolution 
IUE spectra SWP45670 and LWP23967 exposed on 17/09/92 
(i.e. 210 days after the maximum) and the optical/near-IR 
spectra published by \citet{a+96}. As the latter spectroscopy 
was not simultaneous with the former one, we interpolated 
spectra made at 40 and 400 days to 210 days after the maximum. 
It was possible, because the continuum profile did not change 
significantly \citep[cf. Figs.~4 and 5 of][]{a+96}. 
Then we scaled fluxes from the near-UV to those of the IUE 
observations and selected 12 flux-points in the continuum 
between 1280 and 8000\,\AA, considering the influence of 
the iron curtain as proposed by \citet{sa93}. Following to S05 
we modeled the continuum fluxes by a two-component-model 
including contributions from a star and nebula. In our model 
SED (Fig.~6) the source of the stellar radiation has 
an effective radius 
   $\theta_{\rm h} = 2.4\pm0.8\times 10^{-11}$ 
and radiates at 
   $T_{\rm h}$ = 31\,000$\pm$5\,000\,K, 
while the nebula radiates at 
   $T_{\rm e}$ = 44\,000$\pm$5\,000\,K
having the emission measure, 
   $EM = 4\pi d^{2}\times (13.8\pm3)\,10^{15}$\cmt.
We note that the observed stellar radiation is not capable 
of giving rise the nebular emission by photoionization, because 
of too low $T_{\rm h}$. This implies that a major part of 
the nebular radiation is due to collisions, which is consistent 
with its high electron temperature and the presence of a fast 
outflow from the nova. 

The line spectrum of V1974\,Cyg was reconstructed according 
to observations of \citet{ch+93} taken at 212 days (see their 
Figs.~5 and 6) and \citet{a+96} (211 days, Table~7), obtained 
nearly simultaneously with the ultraviolet observations. 
The $U,~B,~V$ measurements obtained on 17/09/92 were taken 
from \citet{ch+93}. 

%
For V1974\,Cyg 
the effect of emission lines on the $U$, $B$, $V$ and $R$ 
magnitudes is significant: 
$\Delta U$ = 1.57, 
$\Delta B$ = 1.73, 
$\Delta V$ = 0.66, 
$\Delta R_{\rm C}$ = 0.94 
and 
$\Delta R_{\rm J}$ = 0.88\,mag. 
Disagreement between the corrected $V$ magnitude and the 
predicted continuum at 5500\,\AA\ is caused by a large scatter 
in the $V$ measurements obtained at different observatories at 
the time of the presence of strong nebular emissions in 
the V1974\,Cyg spectrum \citep[see Fig.~8 of][]{ch+93}. 
Note that the strongest nebular lines 
($\lambda$4959 and $\lambda$5007) are placed at the steep 
edge of the $V$ filter (Fig.~1). 

\section{Discussion}

\subsection{Comparison of exact and simplified solution}

Deviations between $\Delta m$ corrections obtained by 
the accurate calculation (Eq.~2) and by our approximate 
relation (Eq.~6) are less than 10\% for 
$B,~V,~R_{\rm C},~R_{\rm J}$ measurements (Table~2). 
A systematic shift for the $R_{\rm J}$ passband 
($\Delta R_{\rm J}(2) < \Delta R_{\rm J}(6)$) 
is due to the slope of the red giant continuum here. 
This increases the effective wavelength of 
the $R_{\rm J}$ filter and thus the contribution from the 
continuum, which results in a smaller (correct) $\Delta m$ 
than that estimated for a constant continuum (cf. Eq.~(2)). 
Larger differences are for $\Delta U$ corrections, because 
of a marked difference between the true and the assumed 
constant continuum profile around the Balmer discontinuity. 
%
%
To get a satisfactory result in $U$ for objects 
with a strong Balmer jump in emission, we multiplied the sum 
in Eq~(6) by a factor of 0.8. 

\subsection{Examples for practical applications}

Corrections $\Delta m$ can be used to convert arbitrary 
flux units of an emission-line spectrum to fluxes in 
absolute units with the aid of the simultaneous $UBVR$ 
photometry. The following steps are relevant. 

(i) Appropriate $\Delta m$ can be estimated according to 
Eq.~(6) to obtain magnitudes of the line-removed continuum. 

(ii) To deredden the continuum-magnitudes and convert them 
to fluxes as referred in Sect.~3. 
%

(iii) A polynomial fit to the line corrected flux-points 
of the photometric passbands can be adopted as the true 
continuum. 
%

In special cases it is sufficient to limit this calibration 
just to one passband to get an appropriate scaling factor. 
Usually this takes effect for a spectrum exposed within the $R$ 
band which contains one strongly dominant emission from 
\ha\ (e.g. symbiotic stars). In this case Eq.~(6) can 
be expressed as
%
\begin{equation}
\Delta m_{\alpha} = 
  -2.5\log [1 + P_{\rm R}(6563)F_{\alpha}/C_{\rm R}],
\end{equation}
%
where $F_{\alpha}$ is the \ha\ flux in units of the local 
continuum, transmissivities $P_{\rm R_C}(6563)$ = 0.79, 
$P_{\rm R_J}(6563)$ = 0.92 and $C_{\rm R}$ quantities 
(Eq.~(7)) are in Table~1. 

A proper corrections for emission lines must be applied 
to model multicolour light curves of the outbursting objects, 
spectra of which are very rich for emission lines. 
For example, \citet{hachisu+06} restricted their calculation 
of theoretical light curves of the 2006-outburst of 
the recurrent symbiotic nova RS\,Oph only to the $y$ and 
$I_{\rm C}$ filters to avoid contamination by the strong 
emission lines, although $B$ and $R_{\rm C}$ measurements 
were also available. 

The effect of emission lines on the photometric measurements 
changes also the colour indices. Generally, $\Delta m$ 
corrections are different in different passbands, which 
results in a relevant change in the colour indices of 
the true continuum. 
Therefore, prior to a diagnostic of the emission-line objects 
by colour diagrams one has to correct the observed indices 
for the light excess due to the emission lines. 
This approach can be effectively used to study evolution 
of the composite spectra of, for example, symbiotic stars 
on the basis of their multicolour light curves. 
However, this represents rather sophisticated task and 
we will devote to this problem in a separate paper. 

\section{Conclusions}

We investigated the effect on the $U,~B,~V$ and $R$ magnitudes 
of the removal of the emission lines from the spectra of some 
classical symbiotic stars, symbiotic novae and the classical 
nova V1974\,Cyg. Our results may be summarized as follows: 

(i)
We calculated the ratio of fluxes with and without the lines, 
transmitted through the given photometric filter, to obtain 
corrections $\Delta U$, $\Delta B$, $\Delta V$, $\Delta R_{\rm C}$ 
and $\Delta R_{\rm J}$ caused by emission lines. 
First, we approached this problem by a precise reconstruction 
of the true continuum (Eq.~2). Second, we derived a formula to 
estimate these corrections by a simple way for an operative use 
(Eq.~6). Deviations between $\Delta m$ corrections determined 
by both the methods are less than 10\%. Larger differences are 
in the $U$ passband, because of a complex profile of the true 
continuum at this region. 

(ii)
The removal of emission lines makes the star's brightness 
fainter by $\Delta U \sim 0.33$, 
           $\Delta B \sim 0.46$, 
           $\Delta V \sim 0.12$,
           $\Delta R_{\rm C} \sim 0.33$ and 
           $\Delta R_{\rm J} \sim 0.25$\,mag 
for the selected symbiotic stars (Table~2). 
The effect is larger for AX\,Per at the eclipse, because of 
the low level of the continuum (e.g. $\Delta U = 0.77$ and 
$\Delta B = 0.62$\,mag). 
The largest corrections were found for the symbiotic novae 
and the nebular phase of the classical nova V1974\,Cyg 
          ($\Delta U \sim 1.5$,
           $\Delta B \sim 1.7$,
           $\Delta V \sim 0.91$,
           $\Delta R_{\rm C} \sim 1.3$ and
           $\Delta R_{\rm J} \sim 1.1$\,mag), 
because of their very intense emission line spectrum.
The significant effect in $R$ passbands is mainly due to a strong 
\ha\ emission and the high transmissivity of both $R_{\rm C}$ and 
$R_{\rm J}$ filters at $\lambda\,6563$\,\AA\ (Eq.~(10), Table~1, 
Fig.~1). 

(iii)
The line corrected $UBVR$ flux-points fit well the modeled 
continuum (Figs.~2 and 4--6). Thus they can be used in many 
astrophysical applications for studying the continuum of 
emission-line objects (Sect.~4.2). 

\section*{Acknowledgments}

This research has been in part supported by the Slovak Academy 
of Sciences under a grant No.~2/7010/7. 
The code for calculating $\Delta m$ corrections according to 
Eq.~(6) is available at 
   http://www.astro.sk/\,$\tilde {}$\,astrskop/ubvr.corr/ .


\begin{thebibliography}{}

\bibitem[Andrews (1974)]{a74}
         Andrews, P.J., 1974. MNRAS 167, 635

\bibitem[Austin et al. (1996)]{a+96}
         Austin, S.J., Wagner, R.M., Starrfield, S., et al., 
         1996. AJ 111, 869

\bibitem[Belyakina (1992)]{bel92}
         Belyakina, T.S., 1992. Izv. Kr. Astrophys. Obs. 84, 49

\bibitem[Bessell (1979)]{b79}
         Bessell, M.S., 1979. PASP 91, 589
%

\bibitem[Bruch et al. (1994)]{bruch}
         Bruch, A., Niehues, M., Jones, A.F., 
         1994. A\&A 287, 829

\bibitem[Chochol et al. (1993)]{ch+93}
         Chochol, D., Hric, L., Urban, Z., et al.,
         1993. A\&A 277, 103

\bibitem[Collins (1992)]{c92}
         Collins, P., 1992. IAU Circ. No. 5454

\bibitem[Contini and Formiggini (1999)]{cf99}
         Contini, M., Formiggini, L., 1999. ApJ 517, 925

\bibitem[Fern\'andez-Castro et al. (1995)]{f-c+95}
         Fern\'andez-Castro, T., Gonz\'alez-Riestra, R., 
         Cassatella, A., et al.,
         1995. ApJ 442, 366

\bibitem[Fernie (1985)]{f85}
         Fernie, D.J., 1985. PASP 97, 653

\bibitem[Formiggini and Leibowitz (1994)]{fl94}
         Formiggini, L., Leibowitz, E.M., 
         1994. A\&A 292, 534 

\bibitem[Hachisu et al. (2006)]{hachisu+06}
         Hachisu, I., Kato, M., Kiyota, S., et al.,
         2006. ApJ 651, L141

\bibitem[Hauschildt et al. (1999)]{h+99}
         Hauschildt, P.H., Allard, F., Ferguson, J., et al., 
         1999. ApJ 525, 871

\bibitem[Henden and Kaitchuck (1982)]{hk82}
         Henden, A.A., Kaitchuck, R.H., 1982. 
         Astronomical Pho\-to\-metry, Van Nostrand Reinhold 
         Company, New York, 50 

\bibitem[Johnson (1965)]{johnson}
         Johnson, H.L., 1965. ApJ 141, 923

\bibitem[Kenyon et al. (1993)]{k+93}
         Kenyon, S.J., Mikolajewska, J., Mikolajewski, M., et al., 
         1993. AJ 106, 1573

\bibitem[Kotnik-Karuza et al. (2006)]{kk+06}
         Kotnik-Karuza, D., Friedjung, M., Whitelock P. A., et al., 
         2006. A\&A 452, 503

\bibitem[Mao-Lin and Bloch (1954)]{lb54}
         Mao-Lin, T., Bloch, M., 1954. Ann. d' Astrophys. 17, 6

\bibitem[Mao-Lin and Bloch (1957)]{lb57}
         Mao-Lin, T., Bloch, M., 1957. Ann. d' Astrophys. 20, 86

\bibitem[Matthews and Sandage (1963)]{ms63}
         Matthews, T.A., Sandage, A.R., 1963. ApJ 138, 30

\bibitem[Menzies et al. (1982)]{m+82}
         Menzies, J.W., Coulson, I.M., Caldwell, J.R.A.,
         Corben, P.M., 1982. MNRAS 200, 463.

\bibitem[Mikolajewska and Kenyon (1992)]{mk92}
         Mikolajewska, J., Kenyon, S.J., 1992. AJ 103, 579

\bibitem[Mikolajewska and Kenyon (1996)]{mk96}
         Mikolajewska, J., Kenyon, S.J., 1996. AJ 112, 1659

\bibitem[Munari et al. (1992)]{m+92}
         Munari, U., Yudin, B.F., Taranova, O.G., et al., 
         1992. A\&AS 93, 383

\bibitem[Parimucha et al. (2000)]{par+00}
         Parimucha, \v{S}, Arkhipova, V.P., Chochol, D., et al.,
         2000. Contrib. Astron. Obs. Skalnat\'e Pleso 30, 99

\bibitem[Penston et al. (1983)]{p+83}
         Penston, M.V., Benvenuti, P., Cassatella, A., et al.,
         1983. MNRAS 202, 833

\bibitem[Sandage and Eggen (1959)]{seg59}
         Sandage, A.R., Eggen, O.J., 1959. MNRAS 119, 298

\bibitem[Schmeer (1992)]{sch92}
         Schmeer, P. 1992, IAU Circ. No. 5455

\bibitem[Schmid and Schild (1990)]{ss90}
        Schmid, H.M., Schild, H., 1990. MNRAS 246, 84

\bibitem[Shore and Aufdenberg (1993)]{sa93}
         Shore, S.N., Aufdenberg, J.P., 1993. ApJ 416, 355

\bibitem[Skopal (2003)]{sk03}
         Skopal, A., 2003. Baltic Astronomy 12, 604

\bibitem[Skopal (2005)]{sk05}
         Skopal, A., 2005. A\&A 440, 995 (S05)

\bibitem[Skopal (2006)]{sk06}
         Skopal, A., 2006. A\&A 457, 1003

\bibitem[Skopal et al. (2001)]{sk+01}
         Skopal, A., Teodorani, M., Errico, L., et al., 
         2001. A\&A 367, 199

\bibitem[Swings and Struve (1940)]{ss40}
         Swings, P., Struve, O., 1940. ApJ 91, 546

\bibitem[Taranova and Shenavrin (2000)]{ts00}
         Taranova, O.G., Shenavrin, V.I., 
         2000. Astron. Letters 26, 600

\bibitem[Thackeray and Hutchings (1974)]{th74}
         Thackeray, A.D., Hutchings, J.B., 
         1974. MNRAS 167, 319

\bibitem[Tomov and Tomova (1998)]{tt98}
         Tomov N.A., Tomova, M.T., 1998. IBVS No. 4574

\bibitem[Tomov et al. (2003)]{ttt03}
         Tomov, N.A., Taranova, O.G., Tomova, M.T., 2003. 
         A\&A 401, 669

\bibitem[Young et al. (2005)]{y+05}
         Young, P.R., Berrington, K.A., Lobel, A., 
         2005. A\&A 432, 665

\end{thebibliography}
\end{document}